\documentclass[aps,pra,amsmath,amssymb,showpacs,twocolumn]{revtex4-1}

\usepackage[T1]{fontenc}
\usepackage{amssymb,amsmath}
\usepackage{amsfonts}
\usepackage{graphicx}
\usepackage{amsthm}
\usepackage{color}
\usepackage{mathbbol}
\usepackage{epstopdf}

\usepackage{tikz}
\usepackage{verbatim}

\def\vec#1{\mathbf{#1}}
\def\ket#1{|#1\rangle}

\def\bra#1{\langle#1|}
\def\ketbra#1{|#1\rangle\langle#1|}

\def\tr{\mathrm{tr}}

\def\ccs{\textrm{ccs}}

\newcommand{\bn}{{\mathbf n}}

\newenvironment{rcases}
  {\left.\begin{aligned}}
  {\end{aligned}\right\rbrace}

\begin{document}

\title{Quantumness of spin-1 states}
\author{Fabian Bohnet-Waldraff$^{1,2}$, D.~Braun$^{1}$, and O.~Giraud$^{2}$}
\affiliation{$^{1}$Institut f\"ur theoretische Physik, Universit\"at T\"{u}bingen, 72076 T\"ubingen, Germany\\
$^{2}$\mbox{LPTMS, CNRS, Univ. Paris-Sud, Universit\'e Paris-Saclay, 91405 Orsay, France}
}
\begin{abstract}
We investigate quantumness of spin-1 states, defined as the Hilbert-Schmidt distance to the convex hull of spin coherent states. We derive its analytic expression in the case of pure states as a function of the smallest eigenvalue of the Bloch matrix and give explicitly the closest classical state for an arbitrary pure state. Numerical evidence is provided that the exact formula for pure states  provides an upper bound on the quantumness of mixed states. Due to the connection between quantumness and entanglement we obtain new insights into the geometry of symmetric entangled states.
\end{abstract}
\date{\today}
\maketitle

\section{Introduction}

The quantum world is the realm of the most counter-intuitive
phenomena, from the tunnel effect to the more recent quantum
teleportation. There are, however, instances of quantum states which
behave in an almost classical way. The best-known example of such a
behavior is that of coherent states. 
With the rise of quantum information technology the need to identify
genuine quantum states, where truly quantum phenomena could occur, has become
important. Several notions of ``quantumness'' exist, emphasizing
different physical consequences of quantum behaviour.  One of the
oldest ones goes back to quantum optics, where coherent states of
light are considered the most classical pure states possible. These are
states with minimal quantum uncertainty in the quadratures,
i.e.~localized as much as possible in phase space, and this property
is preserved under the free time evolution of the
electro-magnetic field \cite{Schroedinger}.
The purely classical procedure of randomly choosing such states adds
classical noise but no quantum noise.  The resulting mixed states, whose Glauber representation is a convex sum of coherent state density matrices, form
a convex set of states with positive $P$-function,
and there is widespread agreement that such states are to be considered the most
classical states \cite{Mandel86,kim_nonclassicality_2005}. 

This definition was extended to finite-dimensional systems in
\cite{GirBraBra08}, where spin-coherent states (SU(2)
coherent states) play the role of the pure states
with  
minimal quantum fluctuations of the angular-momentum
operators  \cite{Agarwal81}.  
This property is conserved under unitary operations representing
rotations.
A mixed state can be considered classical if it
can be written as a statistical mixture of spin coherent states, meaning
that a representation with a positive $P$-function exists.  The set of 
``classical spin states'' can thus be defined as the convex hull of spin
coherent states
\cite{hillery_nonclassical_1987,GirBraBra08}.
Any state outside this set may be considered truly quantum. To measure
the departure from the classical behaviour it is convenient to
define ``quantumness'' 
as the Hilbert-Schmidt distance from the
state to the set of classical states \cite{QQQ,hillery_nonclassical_1987}. 
Other quantifiers of quantumness are based on different sets of
``classical states'', e.g.~states with positive Wigner function \cite{Kenfack04}, and use various measures of distance, such as the trace distance \cite{Eisert03} or the Bures
distance \cite{Martin10}.  \\

Alternative measures of quantumness are based on entanglement \cite{mari_positive_2012,Bruss01,4Horo}.
Even though formal analogies of entanglement can be
found also in classical physics, and have attracted attention recently in
optics \cite{Collins02}, entanglement is a signature of a quantum behaviour. 
Entangled quantum states can lead to stronger-than-classical
correlations between subsystems. 
A number of entanglement measures have been proposed in order to
quantify entanglement. A way of defining such a measure is to consider
the distance between a state and the convex set of separable states. 
While this
distance 
was shown to yield a good measure of entanglement when it is taken
as the relative entropy or the Bures distance~\cite{VedPleRipKni97,VedPle98}, it is 
currently still unclear whether the Hilbert-Schmidt distance yields a good
measure of entanglement \cite{PRA89-042305-2014}, as it is not contractive \cite{Oza00}.
However, this measure is mathematically convenient  as a Euclidean distance
on Hilbert space,   and has nice 
physical properties.   For 
instance the Hilbert-Schmidt distance 
is equal to the maximum amount by which a certain type of a generalized
Bell inequality is violated \cite{GeoPicOfEnt}.  Furthermore, we show
here that the Hilbert-Schmidt distance gives new insight into the
geometry of entangled states.  \\


In the present paper, we investigate the problem of finding the distance from a state to the set of classical states, as well as the classical state closest to a given state. The closely related problem of finding the {\em separable} state closest to a given state has already been investigated in the literature. For instance, if one restricts the set of separable states to pure states then it was shown in \cite{HubKle09} that the closest separable pure state in terms of Bures distance to a pure symmetric state is always symmetric. This result also holds for the Hilbert-Schmidt distance as both distances are simply related to the overlap of the two states in this pure state case. In \cite{GeometryOfEntStates}, the problem of the Hilbert-Schmidt distance from a bipartite two qubit state to the closest (possibly mixed) separable state was investigated. Specializing the results of \cite{GeometryOfEntStates} to symmetric states, one can observe that the separable state closest to a symmetric state (pure or not) is in general mixed and not necessarily symmetric.

Here we solve the problem of finding the classical state closest to a general \mbox{spin-1} state, in terms of the Hilbert-Schmidt distance. We find an analytical solution for pure states. Our findings generalize a result obtained in \cite{QQQ} for the most quantum \mbox{spin-1} pure state. As we will see, this also solves the problem of finding the symmetric separable state closest to a pure symmetric bipartite state of two qubits.

The paper is organized as follows. In Section \ref{definitions} we introduce the Bloch matrix representation that we will use throughout the paper. Section \ref{purespin1} solves the problem of finding the classical state closest to any given pure \mbox{spin-1} state, while Section \ref{mixed} tackles the problem for mixed states. Section \ref{entanglement} makes the connection with entanglement and entanglement measures.


%

\section{Definitions}
\label{definitions}
\subsection{Tensor representation}\label{sec.tenrep}
A way of representing spin-$j$ states which is particularly convenient when dealing with spin coherent states is the tensor representation proposed in \cite{PRL2015}. It is a generalization of the well-known Bloch picture for spin-$\frac12$ states. In the case $j=\frac12$, any state $\rho$ can be expanded as 
\begin{equation}
\label{bloch12}
\rho=\frac{1}{2} \sum_{\mu=0}^3 X_\mu S_\mu,
\end{equation}
with $S_0$ the $2 \times 2$ identity matrix, and $S_i=\sigma_i$, $1\leq i\leq3$, the three Pauli matrices. In this basis, the coordinates of $\rho$ are $X_0=1$ and $X_i=\tr(\rho S_i)$, so that $\vec{X}=(X_1,X_2,X_3)$ forms the usual Bloch vector. 

For higher spin, it is possible to associate to any spin-$j$ state $\rho$ a tensor with $2j$ indices \cite{PRL2015}. For spin-1 this tensor reduces to a matrix that can be defined as 
\begin{equation}
\label{Bloch matrix}
X_{\mu \nu}=\tr (\rho \, S_{\mu \nu}),\qquad 0\leq \mu,\nu\le 3,
\end{equation}
with $S_{00}=\mathbb{1}$, the $3 \times 3$ identity matrix, $S_{a0}=S_{0a}=J_a$, and
$S_{ab}=J_aJ_b+J_bJ_a-\delta_{ab}\mathbb{1}$, where $J_a$ is the usual spin-1 angular momentum operator, $1\leq a,b\leq 3$ (here we take $\hbar=1$). The matrices $S_{\mu\nu}$ are such that
$\rho$ can be expanded as 
\begin{equation}
\label{Connection Bloch matrix to the state}
\rho=\frac{1}{4} \sum_{\mu,\nu=0}^{3}X_{\mu \nu} S_{\mu \nu}.
\end{equation}
The $4\times 4$ matrix $X$ is real and symmetric with trace two. As in
the spin-$\frac12$ case, where the Bloch vector transforms as a
three-dimensional vector under rotations of the coordinate frame, for
spin-1, $X$ transforms under a 3D rotation according to
$X'=RXR^\dagger$, with $R_{ab}$, $1\leq a,b\leq3$, the $3 \times 3$
rotation matrix, and $R_{0 \mu}=R_{\mu 0}=\delta_{\mu 0}$, $0\leq
\mu\leq3$. We will thus call $X$ the Bloch matrix. 

This representation is particularly well-suited to our problem, since,
as we will see, coherent states take a very simple form in this
framework.

\subsection{Quantumness}
\label{section quantumness}
The set $\mathcal{C}$ of classical spin states is defined \cite{GirBraBra08} 
as the ensemble
of all density matrices which can be expressed as a mixture of 
spin coherent states with positive weights, i.e.~states
$\rho_c$ for which there exist weights $w_i\geq 0$ and coherent states
$\ket{\alpha_i}$ such that 
\begin{equation}
\label{classicalstates}
\rho_c= \sum_i w_i \ket{\alpha_i}\bra{\alpha_i},
\end{equation}
with $0 \leq w_i \leq 1$, and $\sum_i w_i =1$. Here we use the
following definition of spin coherent states
$\ket{\alpha}=\ket{\theta,\phi}$, with $\theta \in [0,\pi]$ and
$\phi \in [0,2\pi[$ the usual spherical angles,
\begin{equation}
\label{spin coherent}
\ket{\alpha} =\!\!\! \sum_{m=-j}^j \sqrt{\binom{2j}{j+m}} \left(\cos\frac{\theta}{2}\right)^{j+m}\left(\sin\frac{\theta}{2}e^{-i\phi}\right)^{j-m}\!\!\!\!\!\! \ket{j,m},
\end{equation}
where $\ket{j,m}$ is the usual spin basis, here with $j=1$ and $m=-1,0,1$. 

The Bloch matrix of a coherent state takes the simple form $X_{\mu\nu}=n_{\mu}n_{\nu}$, $0\leq \mu,\nu\leq 3$, with $n_0=1$ and $\bn=(n_1,n_2,n_3)=(\sin{\theta} \cos{\phi},\sin{\theta} \sin{\phi},\cos{\theta})$. The decomposition \eqref{classicalstates} can be reexpressed in terms of the Bloch matrix $W$ of $\rho_c$ as
\begin{equation}
\label{Bloch matrix classical state}
W_{\mu \nu}=\tr (\rho_c S_{\mu\nu})=\sum_i w_i n^{(i)}_{\mu}n^{(i)}_{\nu},
\end{equation}
with $\bn^{(i)}=(\sin{\theta_i} \cos{\phi_i},\sin{\theta_i} \sin{\phi_i},\cos{\theta_i})$ the Bloch vectors corresponding to coherent states $\ket{\alpha_i}$ and $n^{(i)}_{0}=1$.

Quantumness of an arbitrary state $\rho$ can be defined \cite{QQQ} as the (Hilbert-Schmidt) distance to the convex set $\mathcal{C}$. Namely, the quantumness $Q(\rho)$ is given by  
\begin{equation}
 \label{quantumness}
Q(\rho)=\min_{\rho_c\in\mathcal{C}} ||\rho-\rho_c||,
\end{equation}
where  $||A|| = \sqrt{\tr (A^\dagger A)}$ is the Hilbert-Schmidt
norm. Using Eq.~\eqref{Connection Bloch matrix to the state}, one can
show that the quantumness can be re-expressed in terms of Bloch matrices as
\begin{equation}
\label{quantumnessXW}
Q(\rho)=\frac{1}{2} \min_{W \textrm{classical} } ||X-W||,
\end{equation}
where $X$ is the Bloch matrix of $\rho$ and $W$ is given by \eqref{Bloch matrix classical state}.

In \cite{GirBraBra08}, a necessary and sufficient criterion for
classicality in the spin-1 case was obtained. A spin-$1$ state is
classical if and only if the $3\times 3$ matrix $Z$ defined (using the
present notation) by $Z_{ab}=X_{ab}-X_{a0}X_{b0}$, with $1\le a,b \le 3$,
is positive semi-definite. Remarkably, the matrix $Z$ is nothing but
the Schur complement of the $1\times 1$ upper left block of matrix $X$
(note that $X_{00}=1$). Therefore positive semi-definiteness of $Z$ is
equivalent to positive semi-definiteness of $X$. In other words, a
spin-1 state is classical if and only if its matrix $X$ is positive
semi-definite. 
Equivalently, a spin-1 state is quantum if and only if the smallest eigenvalue of its matrix $X$ is negative.

The Bloch matrix thus provides a simple classicality criterion. In the
case of pure states, it also allows one to obtain an exact expression for the quantumness \eqref{quantumness}. This is the goal of the next section.

\section{Pure States}
\label{purespin1}

Starting from a one-dimensional parametrization of pure states, we now
prove a lower bound to the minimization problem \eqref{quantumness}
and then show that this lower bound can be reached by a classical
state. This gives an analytic expression for $Q$ for all pure states.

\subsection{Parametrization}
The Majorana representation \cite{Majo32,MajoranaExplained} allows one to uniquely map any pure
spin-$j$ state to $2j$ points on the Bloch sphere. If the pure state
undergoes a unitary transformation $e^{i \varphi \bf{J}.\bf{n}}$ that
represents a rotation of angle $\varphi$ about vector $\bf{n}$ then
the Majorana points are rotated rigidly by that rotation. Under such a
transformation, coherent states are
rotated into coherent states, so that from its definition it is clear
that quantumness is invariant under rotation of the coordinate
system. 
Moreover, since $X$ transforms  under rotations as explained in
Section \ref{sec.tenrep}, its 
eigenvalues  are unchanged under such rotations.

The Majorana representation of a spin-1 pure state $\ket\psi$ just
consists of two points on the unit sphere. These points correspond,
via the stereographic projection $z=\cot\frac{\theta}{2}e^{i\phi}$, to
the roots of the Majorana polynomial
$P(z)=d_1-\sqrt{2}d_0z+d_{-1}z^2$, with $d_m, -1\leq m\leq 1$, the
coefficients of the state $\ket\psi$ in the $\ket{j,m}$ basis. The
sphere (or the spin-1 state) can always be rotated in such a way that
these two Majorana points are brought to a canonical position where
they have spherical coordinates $(\theta,\phi)=(\gamma,0)$ and
$(\pi-\gamma,0)$ without changing quantumness. States with Majorana points at positions $(\gamma,0)$ and $(\pi-\gamma,0)$ are given (up to normalisation $\mathcal{N}$) by 
\begin{equation}
\label{canonicalj1}
\ket{\psi_\gamma}=
\mathcal{N}\left(\ket{1,-1}+\frac{\sqrt{2}}{\sin\gamma}\ket{1,0}+\ket{1,1}\right),
\end{equation}
with $\gamma\in [0,\pi/2]$. We will use this expression as a canonical form for spin-1 pure states. The corresponding Bloch matrix $X$ is given by
\begin{equation}
\label{Xlambda}
X=\left(
\begin{array}{cccc}
 1 & \sqrt{1-\lambda ^2} & 0 & 0 \\
 \sqrt{1-\lambda ^2} & 1 & 0 & 0 \\
 0 & 0 & -\lambda  & 0 \\
 0 & 0 & 0 & \lambda  \\
\end{array}
\right),
\end{equation}
with
\begin{equation}
\lambda=\frac{\sin^2\gamma-1}{\sin^2\gamma+1}.
\end{equation}
The eigenvalues of $X$ are $\pm \lambda$ and $1\pm
\sqrt{1-\lambda^2}$. When $\gamma$ varies in $[0,\pi/2]$, $\lambda$ varies in
$[-1,0]$, so that the smallest eigenvalue (and the only negative one) is $\lambda$. We will use $\lambda$ as the parameter for spin-1 pure states.

According to the criterion of Section \ref{section quantumness}, a
state $\rho$  is classical if and only if $X$ is positive
semi-definite, that is, if and only if $\lambda\geq 0$. For pure states, since $\lambda\in[-1,0]$ this implies that $\lambda=0$. The Bloch matrix \eqref{Xlambda} then corresponds to the Bloch matrix of a coherent state with vector $\bn=(1,0,0)$. Another way of seeing this is to note that $\lambda=0$ is equivalent to $\gamma=\pi/2$, which corresponds to both Majorana points coinciding, i.e.~a coherent state. We recover the known fact that the only classical pure states are coherent states. 

\subsection{Lower bound for the full range}
\label{lower1}
We now show that for an arbitrary pure state $\ket{\psi}$ whose Bloch
matrix $X$ has smallest eigenvalue $\lambda$, quantumness is such that 
\begin{equation}
\label{pure states lower bound}
Q(\ket{\psi} )\geq -\sqrt\frac{3}{8} \lambda.
\end{equation} 
Without loss of generality, the quantumness of 
$\ket{\psi}$ can be calculated by first transforming it
to the canonical form \eqref{canonicalj1}. 
Then we write the quantumness \eqref{quantumnessXW} as
\begin{equation}
\label{quantumness in tensor form}
Q(\ket{\psi})=\frac{1}{2}\min_{W \textrm{classical} }  \sqrt{\sum_{\mu\nu=0}^3 \left(X_{\mu\nu}-W_{\mu\nu}\right)^2},
\end{equation}
with $W$ of the form \eqref{Bloch matrix classical state} and $X$
given by
\eqref{Xlambda}. In order to obtain \eqref{pure states lower bound} it is sufficient to show that $\sum_{\mu\nu}(X_{\mu\nu}-W_{\mu\nu})^2\geq \frac{3}{2}\lambda^2$ for all classical states $W$. 
This is possible by proving: 
\begin{align}
\label{leftrange_a}&(X_{\mu\nu}-W_{\mu\nu})^2 \geq 0,\\
\label{leftrange_b}&(X_{33}-W_{33})^2-\lambda^2\geq 0,\\
\label{leftrange_c}&(X_{11}-W_{11})^2+(X_{22}-W_{22})^2-\frac{\lambda^2}{2}\geq 0. 
\end{align}
The first claim is true for all $\mu,\nu$, since the entries of $X$
and $W$ are real 
numbers. Using \eqref{Xlambda}, condition \eqref{leftrange_b} can be rewritten as
\begin{equation}
\left(|\lambda|+W_{33}\right)^2-\lambda^2 \geq 0,
\end{equation}
which obviously holds since $W_{33}=\sum_i w_i \cos^2\theta_i\geq 0$. In order to prove \eqref{leftrange_c}, we define $a=(W_{11}+W_{22})$ and $b=(W_{11}-W_{22})$. Then one can show the identity
\begin{multline}
(X_{11}-W_{11})^2+(X_{22}-W_{22})^2-\frac{\lambda^2}{2}\\
=\frac{1}{2} \left[ (1-a)^2-2 \lambda (1-a) + (\lambda +1-b)^2\right].
\end{multline}
Noting that $a=\sum_i w_i \sin^2 \theta_i\in [0,1]$ and $\lambda \leq 0 $, it immediately follows that this quantity is non-negative, which completes the proof of \eqref{pure states lower bound}. 

\subsection{Exact value of  $Q(\ket{\psi})$ for $\lambda \in [-1,-\frac12]$}
\label{exact1}
It turns out that in the parameter range  $\lambda \in [-1,-\frac12]$ there is a classical state at precisely the distance given by the lower bound \eqref{pure states lower bound}. We consider the family of states of the form
\begin{eqnarray}
\label{high quantumness ccs}
\rho_c(w,\beta)&=&(1-2w) \ket{\frac{\pi}{2},0}\bra{\frac{\pi}{2},0}\\
&+&w\ket{\frac{\pi}{2},\beta}\bra{\frac{\pi}{2},\beta}+w\ket{\frac{\pi}{2},-\beta}\bra{\frac{\pi}{2},-\beta},\nonumber
\end{eqnarray}
which are classical by construction for $w \in[0,1/2]$, since they are a mixture of coherent states $\ket{\theta,\phi}$. 
By calculating the unconstrained minimum  
\begin{equation}
\min_{w,\beta}||\ketbra{\psi}-\rho_c(w,\beta)||,
\end{equation}
the optimal choices for the parameters $w$ and $\beta$ are found to be
\begin{equation}
w=\frac{(4 \lambda +2)(1- \sqrt{1-\lambda ^2})-\lambda ^2}{17 \lambda +8}
\end{equation}
and 
\begin{equation}
\beta=\arccos\left(\frac{-\sqrt{1-\lambda ^2}-2 \lambda -1}{2 \lambda }\right).
\end{equation}
The condition $w\in [0,1/2]$ translates to $\lambda \leq -1/2$. For these
values the
Bloch matrix of the state \eqref{high quantumness ccs} reduces to 
\begin{equation}
\label{Wlambda}
W=\left(
\begin{array}{cccc}
 1 & \sqrt{1-\lambda ^2} & 0 & 0 \\
 \sqrt{1-\lambda ^2} & 1+\frac{\lambda}{2} & 0 & 0 \\
 0 & 0 & -\frac{\lambda}{2}  & 0 \\
 0 & 0 & 0 & 0 \\
\end{array}
\right).
\end{equation}
Note that since the set of classical states is closed and convex
\cite{GirBraBra08}, there is a unique closest state to any given state
for the (Euclidean) Hilbert-Schmidt distance. Since the distance from
the state $W$ to the state $X$ \eqref{Xlambda} is exactly the value of
the lower bound, it shows that $W$ represents the 
closest classical state ($\ccs$) for $\ket{\psi}$ in the range of
$\lambda \in [-1,-\frac12]$. 

If $\lambda>-\frac12$ the state corresponding to \eqref{Wlambda} does not
represent a quantum state any more, since the corresponding density
matrix is no longer positive. Actually, in the next section we find a
tighter lower bound for $\lambda\in ]-\frac12,0]$, which in particular
implies that  the distance between a quantum state and the set
$\mathcal{C}$ is larger than $\sqrt{3/8}|\lambda|$ for $\lambda\in
]-\frac12,0[$. This proves that no classical
state exists in
this range $\lambda\in
]-\frac12,0[$ that saturates the bound \eqref{pure states lower bound}.

\subsection{Tighter bound in the range $\lambda \in ]-\frac12,0]$}

In this section we show that for $\lambda \in ]-\frac12,0]$ one has
\begin{equation}
\label{lower bound low quantumness}
Q(\ket\psi)\geq \frac12 \sqrt{ \lambda^2 +\ell(\lambda)},
\end{equation}
where $\ell(\lambda)$ is given by 
\begin{multline} 
\label{l of lambda explicit definition}
\ell(\lambda)=\frac{1}{216} \left[3 h^5 \sqrt{\frac{1-\lambda }{(\lambda +1)^3}}-\frac{6 h^2 (\lambda^2 -52 \lambda +55)}{\lambda
   +1} \right.\\ 
\left.  +h^4 -216 h \sqrt{1-\lambda ^2}+72 (11-4 \lambda^2 + 4\lambda ) 		\vphantom{\sqrt{\frac{1-\lambda }{(\lambda +1)^3}}}		\right]
\end{multline}
with \begin{align}
h=6^{1/3} \left[9 \sqrt{1-\lambda ^2}+ \sqrt{3(\lambda +1) \left(25- 31\lambda -2 \lambda^2 \right)}\right]^{1/3}.
\end{align}
The bound \eqref{lower bound low quantumness} is tighter than the one obtained in Section \ref{lower1}, as can be shown by proving that over the range $\lambda \in ]-\frac12,0[$ one has
\begin{equation}
\label{l of lambda greater than f}
\frac{\sqrt{\lambda^2+\ell(\lambda)}}{2}>-\sqrt{\frac38}\lambda
\end{equation} 
(see end of the Appendix).

In order to prove the lower bound \eqref{lower bound low quantumness} it is sufficient to show that $\sum_{\mu\nu}(X_{\mu\nu}-W_{\mu\nu})^2\geq \lambda^2 + \ell(\lambda) $ for all classical states. This is possible by proving: 
\begin{align}
\label{rightrange_a}& (X_{\mu\nu}-W_{\mu\nu})^2 \geq 0,\\
\label{rightrange_b} & (X_{33}-W_{33})^2-\lambda^2\geq 0,\\
\label{rightrange_c}&  (X_{11}-W_{11})^2+(X_{22}-W_{22})^2+2(X_{01}-W_{01})^2\geq \ell(\lambda). 
\end{align}
Conditions \eqref{rightrange_a} and \eqref{rightrange_b} were already
proven in the previous section, so we only have to show
\eqref{rightrange_c}. This can be done by analytically calculating the
minimal value of the left-hand side of \eqref{rightrange_c} under the
restrictions on the values of $W_{\mu\nu}$ implied by Eq.~\eqref{Bloch
  matrix classical state}. For readability, we rewrite the left-hand
side of \eqref{rightrange_c}, using the form \eqref{Xlambda} for a
general pure state Bloch matrix $X$ and  Eq.~\eqref{Bloch matrix classical state} for a general classical state $W$ as 
\begin{align}
\label{mingoal}
F(u,v,g):=(1-u)^2+(\lambda+v)^2+2(\sqrt{1-\lambda^2}-g)^2
\end{align}
with $u=\sum_i w_i \sin^2\theta_i \cos^2\phi_i$, $v=\sum_i w_i \sin^2\theta_i \sin^2\phi_i$, and $g=\sum_i w_i \sin\theta_i \cos\phi_i$.
These new variables are such that
\begin{align}
u+v \leq 1 \nonumber \\ \label{para range}
u,v \geq 0\\ \nonumber
-\sqrt{u} \leq g \leq \sqrt{u}.
\end{align}
The last condition is derived from Jensen's inequality $\left(\sum_i w_i a_i\right)^2\leq \sum_i w_i a_i^2$ with $a_i= \sin\theta_i \cos\phi_i$.  
The minimum of $F(u,v,g)$ under the constraints \eqref{para range} can be calculated analytically, and, as shown in the Appendix, it is equal to $\ell(\lambda)$ given in \eqref{l of lambda explicit definition}. This proves Eq.~\eqref{rightrange_c}, and thereby the tighter lower bound \eqref{lower bound low quantumness} for the range $\lambda \in ]-\frac12,0]$.  

\subsection{Exact value of $Q(\ket{\psi})$ for $\lambda \in ]-\frac12,0]$}
The tighter lower bound \eqref{lower bound low quantumness} can be
reached in the range of $\lambda \in ]-\frac12,0]$, since there are classical states at this distance. Using a similar approach as in Section \ref{exact1}, we consider a family of classical states of the form 
\begin{equation}
\label{rhoc2}
\rho_c(\beta)=\frac12\left( \ket{\frac{\pi}{2},\beta}\bra{\frac{\pi}{2},\beta}+ \ket{\frac{\pi}{2},-\beta}\bra{\frac{\pi}{2},-\beta}\right),
\end{equation}
which are a mixture of just two coherent states $\ket{\theta,\phi}$
with equal weights $\frac12$. Let a pure state $\ket{\psi}$ have a
Bloch matrix with smallest eigenvalue $\lambda$. The state
$\rho_c(\beta)$ closest to the
canonical form (\ref{canonicalj1}) of $\ket{\psi}$ is determined by the condition   
\begin{equation}
\frac{\partial}{\partial_\beta} ||\ketbra{\psi}-\rho_c(\beta)||=0,
\end{equation}
which has the solution $\beta=\arccos d$, with $d$ defined as the real root of the polynomial  
\begin{equation}
\label{PolynomThatGivesTheAngle}
P(y)=\sqrt{1-\lambda^2}+y(1+\lambda)-2y^3,
\end{equation}
where $\lambda \in ]-\frac12,0]$, corresponding to $d\in]\frac{\sqrt3}{2},1]$.
Using this value of $\beta$ gives the Bloch matrix of $\rho_c$ as
\begin{equation}
\label{Wlambda2}
W=\left(
\begin{array}{cccc}
 1 & d & 0 & 0 \\
 d & d^2 & 0 & 0 \\
 0 & 0 & 1-d^2 & 0 \\
 0 & 0 & 0 & 0 \\
\end{array}
\right).
\end{equation}
The state represented by \eqref{Wlambda2} is then exactly at the distance to the pure state
\eqref{Xlambda} given by the tighter lower bound \eqref{lower bound low quantumness}. Therefore we have proven that the classical state closest to \eqref{Xlambda} is \eqref{Wlambda2} for the parameter range
$\lambda \in ]-\frac{1}{2},0]$.

\subsection{Summary of results for pure states}
\label{summary}

To conclude, let an arbitrary pure spin-1 state $\ket\psi$ be given by
its Bloch matrix \eqref{Bloch matrix}. If the smallest eigenvalue of
$X$ is denoted by $\lambda$, then the quantumness of $\ket\psi$ is
equal to the quantumness of a state with Bloch matrix \eqref{Xlambda},
and takes the form 
\begin{equation}
\label{analytic quantumness pure states}
Q(\ket{\psi})=f(\lambda),
\end{equation}
with 
\begin{align}
\label{foflambda}
f(\lambda):=\begin{cases}
-\sqrt{\frac38} \lambda &\text{for $\lambda \leq -\frac12$,}\\
\frac12 \sqrt{\lambda^2+\ell(\lambda)} &\text{for $\lambda > -\frac12$,}\\
\end{cases}
\end{align}
and $\ell(\lambda)$ given by Eq.~\eqref{l of lambda explicit definition}. 
The function $f(\lambda)$ is shown in Fig.~\ref{fig:QVsEW}. It is
continuous at $\lambda=-\frac12$. At this plot scale, $f(\lambda)$ is
almost indistinguishable from a linear function. The maximal
difference between $f(\lambda)$ and $-\sqrt{\frac38} \lambda$ is less
than $0.0016$ over the interval $[-1,0]$. 

The classical states closest to a pure state $\ket{\psi}$ take a
different expression in the two 
regions $\lambda<-\frac12$ and $\lambda>-\frac12$, respectively given
by \eqref{Wlambda} and \eqref{Wlambda2}. 
In contrast to the case of the queen of quantum for $j=1$,
corresponding to $\lambda=-1$ \cite{QQQ}, these closest classical
states ($\ccs$) are 
not simply a mixture of the pure state $\ket\psi$ itself 
and the maximally mixed state, i.e. for $\lambda\ne -1$
\begin{equation}
\ccs \neq a \ketbra\psi + (1-a) \frac{\mathbb{1}}{3}, \quad 0\leq a \leq 1.
\end{equation}

\section{Mixed states}
\label{mixed}

\subsubsection{Mixed state quantumness}

For pure states we obtained the analytical expression \eqref{analytic
  quantumness pure states} for quantumness as a function of a single
parameter, namely the smallest eigenvalue  $\lambda$ of the Bloch
matrix of the state. In this Section we investigate the dependence of
$Q(\rho)$ as a function of $\lambda$ for mixed states. For a given
state $\rho$ the quantumness can be obtained numerically by
determining the closest classical state of $\rho$. To find this state
we randomly generate a large sample of coherent states
$\{\ket{\theta_i,\phi_i}\}$ \eqref{spin coherent}, and then optimize
the weights $w_i$ of this decomposition 
\begin{equation}
\label{sommecoh}
\rho_c=\sum_i w_i \ketbra{\theta_i,\phi_i},
\end{equation}
so that the distance from $\rho$ to $\rho_c$ is minimal. As the
function $Q^2$ defined in \eqref{quantumness} corresponds to the
minimization of a function which is quadratic in the $w_i$, this
optimization can be done by quadratic programming. The result of the
optimization yields an approximation of the quantumness: In general it
is overestimated by this approach, as coherent states appearing in the
decompositions of the closest classical state may not be included in
our random sample. However, this overestimated value is very close to
the true value. Indeed, for pure states, where the analytic expression
\eqref{analytic quantumness pure states} is available, the 
error incurred by the numerical approach is typically of the order of
$\sim 10^{-6}$ for the numerical data in this paper. This can be also
checked on mixed states, for instance 
by increasing the number of coherent states involved in the sum
\eqref{sommecoh}, or by changing the choice of coherent states in the
random list: Our results show that the results obtained are quite
independent on increasing the number of coherent states over which we
optimize (further detail on our optimization procedure will be given
elsewhere \cite{Fabian2}).

\begin{figure}[t!]
\begin{center}
\includegraphics[width=0.48\textwidth]{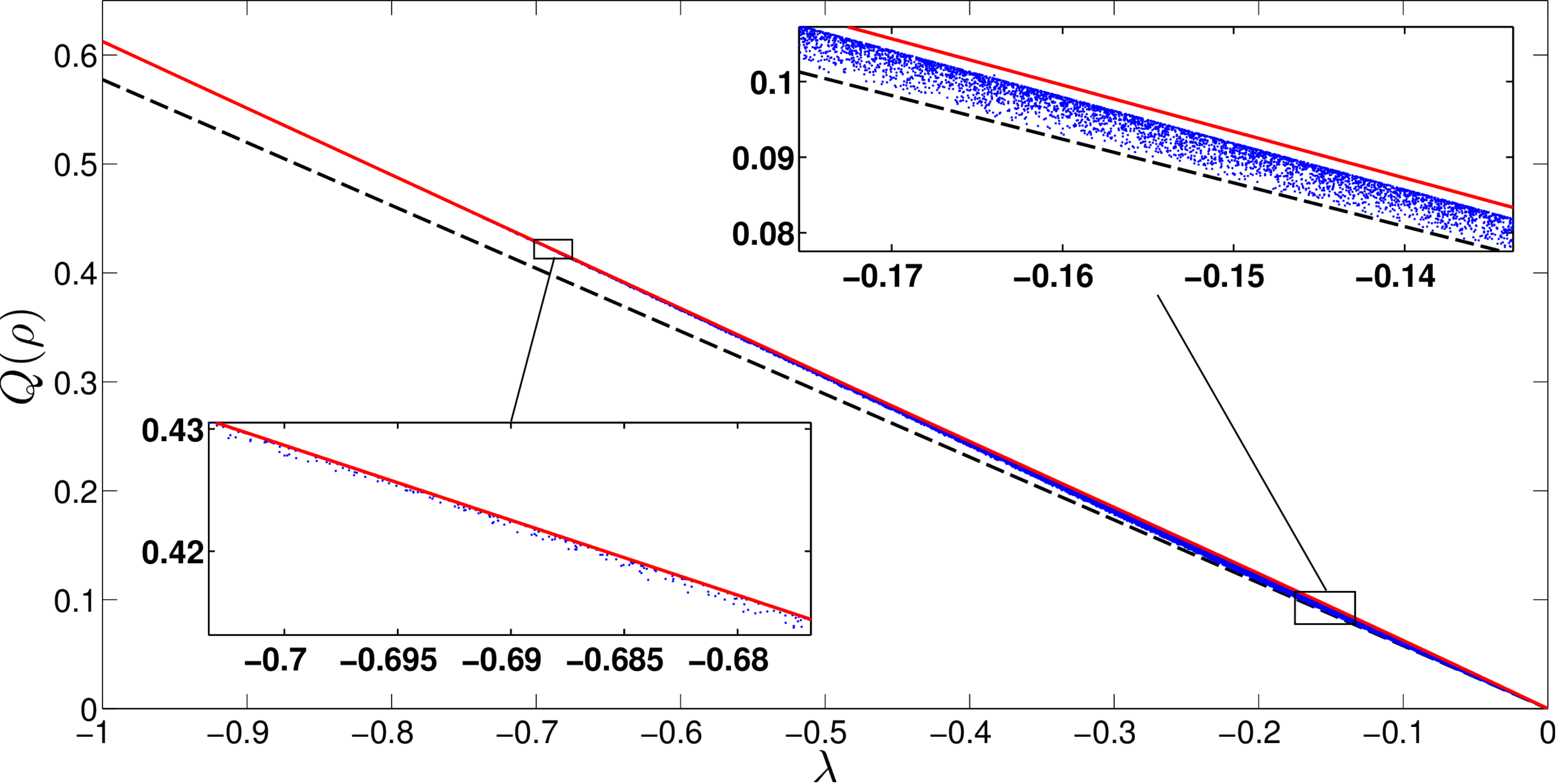}
\end{center}
\caption{(Color online) Quantumness of randomly generated mixed states, as a function of the smallest eigenvalue of their Bloch matrix. There are 50000 points, each one corresponding to a random state. Red line corresponds to the pure states analytic result $f(\lambda)$ given by Eq.~\eqref{foflambda}. Dashed line indicates the lower bound \eqref{lowbound}. Function $f(\lambda)$ appears to be an upper bound on the quantumness of mixed states (see inset).}
\label{fig:QVsEW}
\end{figure}

In Fig.~\ref{fig:QVsEW} we plot the quantumness of mixed states as a
function of the smallest eigenvalue of their Bloch matrix. Since, from
the classicality criterion $X\geq 0$, non-classical states are such
that $\lambda\in[-1,0[$, we restrict our plot to this interval (note
that for classical states $\lambda$ takes positive values, while the
quantumness is zero by definition). The mixed states were randomly
generated from the Hilbert-Schmidt ensemble of matrices
$\rho=AA^{\dagger}/\tr(AA^{\dagger})$, with $A$ a complex matrix with
independent Gaussian entries (see \cite{zyczkowski_generating_2011}
for detail). All points lie very close to the pure state result. This
signifies that quantumness of a mixed state appears to largely depend
on a single parameter, which is the smallest eigenvalue of its Bloch
matrix, although mixed states cannot be reduced to a one-parameter
family (as is the case for pure states, up to rotations).

\subsubsection{Upper bound for quantumness of mixed states}
The function $f(\lambda)$ appears to be an upper bound for the
quantumness of mixed states, namely 
\begin{equation}
Q(\rho)\leq f(\lambda),
\label{upperboundmixed}
\end{equation} 
with $\lambda$ the smallest eigenvalue of the Bloch matrix of
$\rho$. This can be seen in the inset of Fig.~\ref{fig:QVsEW}. More
precisely, Fig.~\ref{fig:QuantMinusBound} displays the difference
between the quantumness and $f(\lambda)$ as a function of the smallest
eigenvalue of the Bloch matrix. In fact, we were not able
to find a single state which violates the bound. It may happen that,
for states very close to pure states, the numerical overestimation of
quantumness due to our optimization procedure leads to a result larger
than $f(\lambda)$; however by increasing the accuracy of our
estimation (that is, taking more coherent states in the sum
\eqref{sommecoh}), we were always able to get this estimate back below
the threshold $f(\lambda)$.   Numerical evidence thus suggests that this
upper bound is valid for all mixed states. 

The almost empty region in the upper right corner of
Fig.~\ref{fig:QuantMinusBound} (visible also just below the upper
bound in the upper inset of
Fig.~\ref{fig:QVsEW}) corresponds to the region between 
$f(\lambda)$ and the straight line $-\sqrt{3/8}\lambda$ in the
interval $[-\frac12.0]$. This apparent emptiness just comes from our
numerical sampling: indeed, this region can be filled e.g.~by points
corresponding to mixed states of the form 
\begin{equation}
\label{mixed case 1}
\rho=a \ketbra{\psi} + (1-a) \ccs(\ket{\psi})
\end{equation}
with $0\leq a\leq 1$ and $\ket{\psi}$ a pure state with closest classical state ccs$(\ket{\psi})$ and $\lambda\in[-\frac12.0]$.


In the special case where a mixed state $\rho$ can be written as a
convex combination of a pure state and its closest classical state
$\ccs$, as in \eqref{mixed case 1}, Eq.~\eqref{upperboundmixed}
can be proven. This can be shown by the fact that
$||\rho-\ccs(\ket{\psi})||=a 
Q(\psi)$, so that $Q(\rho)\leq a f(\lambda_{\psi})$, with
$\lambda_{\psi}$ the smallest eigenvalue of the Bloch matrix of
$\ket\psi$. By using the explicit form of $f$ given by
\eqref{foflambda}, one can show that $a f(\lambda_{\psi})\leq
f(a\lambda_{\psi})$, for $0 \leq a \leq 1$ (this is true for
$-1\leq\lambda\leq -1/2$ because of the inequality \eqref{l of lambda
  greater than f}, and for $-1/2\leq\lambda\leq 0$ by concavity of $f$ 
over this interval). From the forms \eqref{Xlambda} and
\eqref{Wlambda},\eqref{Wlambda2} of the Bloch matrices, one can show 
that for states (\ref{mixed case 1}) the smallest eigenvalue
of the Bloch matrix of $\rho$
is given by
$\lambda=a\lambda_{\psi}$, hence (\ref{upperboundmixed}). 
This proves the upper bound for the family of states \eqref{mixed case 1}. However a proof for arbitrary mixed states is still missing.

\subsubsection{Lower bound for quantumness of mixed states}
The quantumness of mixed states can be bound from below by minimizing over a larger set than in Eq.~\eqref{quantumnessXW} (see \cite{GeometryOfEntStates} for a similar approach). Let $X$ be the Bloch matrix of some state $\rho$ and $\lambda$ be the smallest eigenvalue of $X$. A lower bound can be obtained as
\begin{equation}
\label{NewLowerBoundMixed}
\frac12 \min_{W \text{classical}} ||X -W|| \geq \frac12 \min_{\tilde{W},\tilde{X}} ||\tilde{X} - \tilde{W}||,
\end{equation}
where
$\tilde{W}$ runs over all positive semi-definite matrices with
$\tr\tilde{W}=2$, and $\tilde{X}$ runs over all real symmetric matrices with one eigenvalue equal to $\lambda$ and $\tr \tilde{X}=2$.  
Furthermore,  
we can write $\tilde{X}$ in its diagonal form
$\tilde{X}= \text{diag}(x_1,x_2,x_3,\lambda)$ with $x_i$ arbitrary real
numbers since the norm and the set over which $\tilde{W}$ runs in the
rhs of Eq.~\eqref{NewLowerBoundMixed} are
invariant under orthogonal transformations.

\begin{figure}[t!]
\begin{center}
\includegraphics[width=0.48\textwidth]{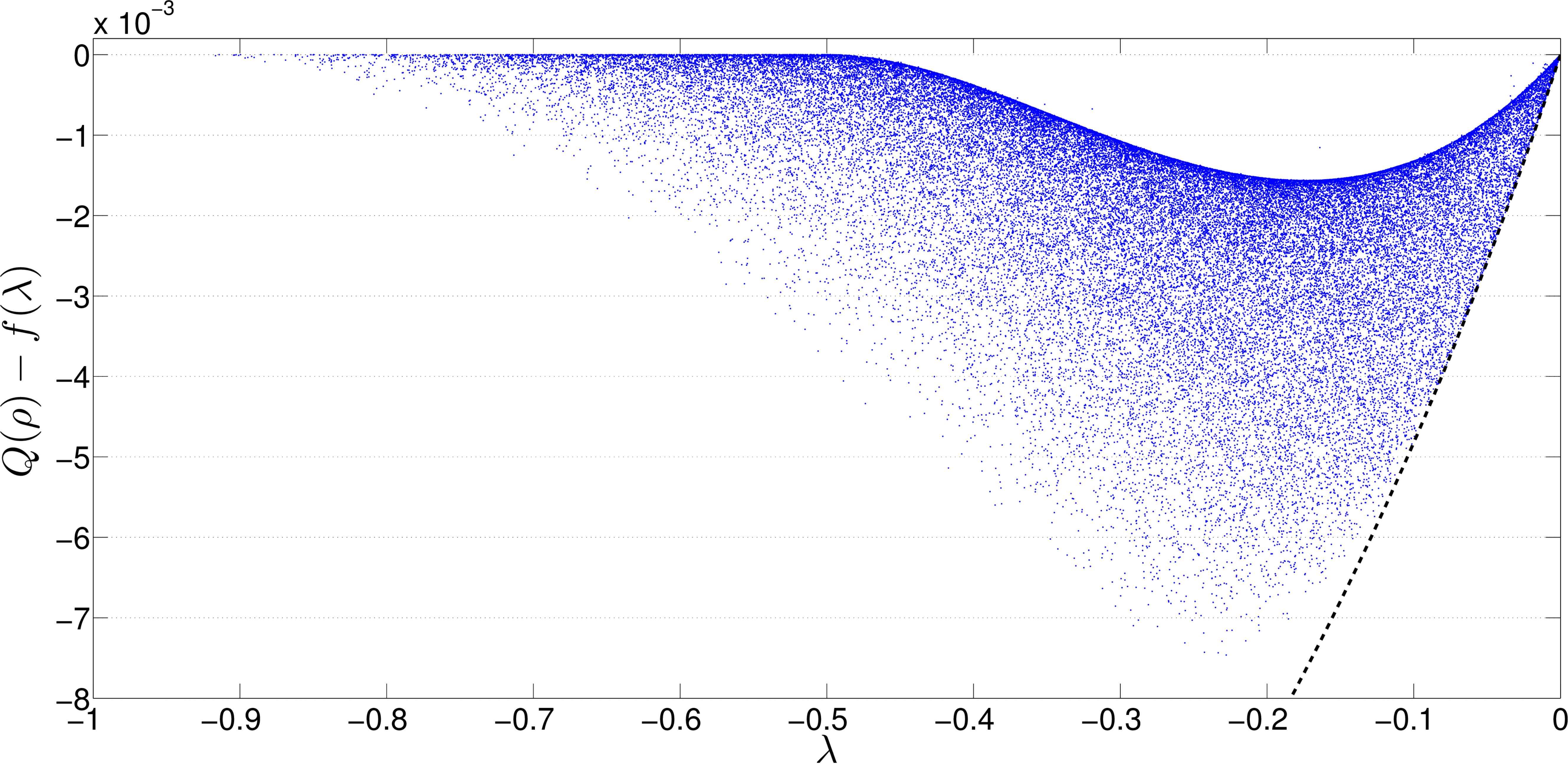}
\end{center}
\caption{Difference between the quantumness and the hypothetical
  upper bound $f(\lambda)$ as function of the smallest eigenvalue of
  the Bloch matrix (same data as in Fig.~\ref{fig:QVsEW}). The difference between the upper bound and the
  quantumness is of the order of $10^{-3}$. The numerical error is of
  order $10^{-6}$, and our numerical procedure
  can only  overestimate quantumness so that the points could only be lower
  than they appear here by that amount. The dashed line corresponds to the lower bound \eqref{lowbound}.} 
\label{fig:QuantMinusBound}
\end{figure}

Because $\tilde{X}$ is diagonal the optimal $\tilde{W}$ will also be in diagonal form. Since $\tilde{W}$ is positive, let $\tilde{W}=\text{diag}(w^2_1,w^2_2,w^2_3,w^2_4)$, with real $w_i$ such that $\sum_{i=1}^4 w_i^2=2$. The right hand side of \eqref{NewLowerBoundMixed} can then be rewritten as
\begin{equation}
\frac12 \min_{\substack{x_i,w_i \in \mathbb{R} \\ \sum_{i=1}^3 x_i =2-\lambda \\ \sum_{i=1}^4 w^2_i=2}} \left[(\lambda-w^2_4)^2 + \sum_{i=1}^3 (x_i-w_i^2)^2    \right]^{1/2}.
\end{equation}
This is a simple problem of minimization under constraints, which can be solved by introducing appropriate Lagrange multipliers. When $\lambda$ is negative (non-classical states), the critical points of the Lagrange function are found to be such that either $w_i=0$ for $1\leq i\leq 3$, or $w_4=0$. The latter case yields the smallest value for the quantumness, which is equal to $-\frac{\lambda}{\sqrt{3}}$. So the quantumness of any mixed state $\rho$ with smallest eigenvalue $\lambda$ of its Bloch matrix is bound by
\begin{equation}
\label{lowbound}
Q(\rho)\geq - \frac{\lambda}{\sqrt{3}}.
\end{equation}
This lower bound corresponds to the dashed line in
Figs.~\ref{fig:QVsEW} and ~\ref{fig:QuantMinusBound}. For small enough quantumness, the two bounds 
provided in this section are close to tight in the sense that the
quantumness of random
mixed states extends almost over the whole range between them. 

\section{Connection with Entanglement}
\label{entanglement}

We now establish a connection between quantumness and entanglement, and relate the smallest eigenvalue of the Bloch matrix to known entanglement measures such as the negativity and the concurrence.

\subsection{Entanglement}
A bipartite pure state $\ket{\psi}$ is called separable if it can be
written as a direct product of pure states of its subsystems
\begin{equation}
\ket\psi=\ket{\psi^{(1)}} \otimes \ket{\psi^{(2)}}.
\end{equation}
This definition can be extended to mixed states: a bipartite mixed
state is called separable if it can be written as a convex sum of
tensor products of quantum states of the subsystems, 
\begin{equation}
\label{sep mixed states}
\rho = \sum_i w_i \rho_i^{(1)} \otimes \rho_i^{(2)},
\end{equation}
where the $w_i$ are classical probabilities with $w_i \geq 0$ and $\sum_i w_i = 1$.
If a state cannot be written in this 
form then it is called entangled \cite{Werner89}.

For  two spin-$\frac12$ states, entanglement can be detected by use of
the partial transpose \cite{Per96}. The necessary and sufficient
'positive partial transpose' (PPT) criterion \cite{Horodecki19961}
states that a state is separable if and only if $\rho^{\textrm{PT}}$ is positive semi-definite, where $^{\textrm{PT}}$ denotes the partial transpose operation. 

In order to quantify entanglement, commonly used measures are the negativity and the concurrence. The negativity is given as  
\begin{equation}
\label{negativity}
\mathcal{N}(\rho) = \sum_i \frac{|\mu_{i}|-\mu_{i}}{2},
\end{equation}
where $\mu_i$ are the eigenvalues of $\rho^{\textrm{PT}}$. The concept of negativity is
also connected to the concept of robustness of entanglement
\cite{RobustnessOfEnt}. 

The concurrence $C$ was developed as an analytic solution of the entanglement of formation for two spins-$\frac12$ \cite{Wooters}. For a two spin-$\frac12$ state $\rho$ it is given as 
\begin{equation}
\label{concurrence}
C(\rho)=\max\{0,\tau_1-\tau_2-\tau_3-\tau_4 \},
\end{equation}
where $\tau_i$ are the square roots of the eigenvalues of the matrix
\begin{equation}
\rho \left(\sigma_y \otimes \sigma_y \right) \rho^* \left(\sigma_y \otimes \sigma_y \right)
\end{equation}
in decreasing order, and $^*$ denotes the complex conjugation. In Section \ref{Quantum-ent} we will relate these entanglement measures with quantumness. We first discuss the analogy between classicality and separability.

\subsection{Classicality and separability}
Classicality is a property defined for a spin-$j$ state. It is
interesting to look at a spin-$j$ state as the projection of a tensor
product of $2j$ spin-$\frac12$ states onto the subspace symmetric
under permutation of the particles. Any
basis vector $\ket{j,m}$ then appears as a symmetrized $2j$-fold tensor
product. 

The symmetric subspace of two spin-$\frac12$ states is spanned by the
Dicke states 
\begin{equation}
\label{dicke}
\ket{D_0}=\ket{\uparrow\uparrow},\quad\ket{D_1}=\frac{1}{\sqrt{2}}(\ket{\uparrow\downarrow}+\ket{\downarrow\uparrow}),\quad\ket{D_2}=\ket{\downarrow\downarrow}.
\end{equation}
The basis vector $\ket{1,m}$ corresponds to $\ket{D_{1-m}}$ for
$-1\leq m\leq 1$. Identifying (in $|j,m\rangle$ notation) $\ket{\frac12,\frac12}=\ket{\uparrow}$ and $\ket{\frac12,-\frac12}=\ket{\downarrow}$, the tensor product of spin-$\frac12$ coherent states \eqref{spin coherent} is 
\begin{multline}
\left(\cos\frac{\theta}{2}\ket{\uparrow}+\sin\frac{\theta}{2}e^{-i\phi}\ket{\downarrow}\right)^{\otimes 2}=\cos^2\frac{\theta}{2}\ket{D_0}+\\
+\sqrt{2}\cos\frac{\theta}{2}\sin\frac{\theta}{2}e^{-i\phi}\ket{D_1}+\sin^2\frac{\theta}{2}e^{-2i\phi}\ket{D_2}
\end{multline}
which, from the correspondence $\ket{1,m}=\ket{D_{1-m}}$, is equivalent to 
\begin{equation}
\label{tensorcoh}
\ket{\alpha}^{j=\frac12} \otimes \ket{\alpha}^{j=\frac12}=\ket{\alpha}^{j=1}
\end{equation}
where $\ket{\alpha}^{j}$ is a spin-$j$ coherent state given by
\eqref{spin coherent}. 
Thus the spin-1 coherent states are separable in the tensor product space. Therefore, all classical states of the form \eqref{classicalstates}, as mixtures of coherent states, can be identified with separable states.

Conversely, a two qubit symmetric separable state $\rho_s=\sum_i w_i \rho_{i}^{(1)} \otimes \rho_{i}^{(2)}$ with $w_i >0$ can be identified with a classical spin-1 state. Indeed, if $\rho_s$ is symmetric then
\begin{equation}
\bra{D^-} \rho_s \ket{D^-}=0,
\end{equation}
with $\ket{D^-}=\frac{1}{\sqrt{2}}(\ket{\uparrow\downarrow}-\ket{\downarrow\uparrow})$.
This is equivalent to
\begin{equation}
\sum_i w_i \bra{D^-} \rho_{i}^{(1)} \otimes \rho_{i}^{(2)} \ket{D^-}=0\,.
\end{equation}
 Since all summands are non-negative (by positivity of density matrices) it follows that
\begin{equation}
\bra{D^-} \rho_{i}^{(1)} \otimes \rho_{i}^{(2)} \ket{D^-}=0 \quad \forall i.
\end{equation}
If the qubit states $\rho^{(1,2)}$ are written with the Bloch vectors
$\vec{X}^{(1,2)}$ according to Eq.~\eqref{bloch12}, direct calculations give
\begin{equation}
\label{blochvector sep symetric}
\bra{D^-} \rho_{i}^{(1)} \otimes \rho_{i}^{(2)} \ket{D^-}=\frac14 (1-\vec{X}^{(1)}\cdot\vec{X}^{(2)}).
\end{equation}
Because the Bloch vectors are such that $||\vec{X}^{(1,2)}||\leq 1$, Eq.~\eqref{blochvector sep symetric} implies that $\vec{X}^{(1)}=\vec{X}^{(2)}$ and $||\vec{X}^{(1,2)}||=1$. Thus $\rho_{i}^{(1)}=\rho_{i}^{(2)}$, which corresponds to the same pure qubit state $\ketbra{\alpha_i}^{j=1/2}$. Therefore one can write
\begin{equation}
\rho_s=\sum_i w_i \ketbra{\alpha_i}^{j=1/2} \otimes \ketbra{\alpha_i}^{j=1/2}.
\end{equation}
With \eqref{tensorcoh} it follows that $\rho_s$ can be identified with 
\begin{equation}
\sum_i w_i \ketbra{\alpha_i}^{j=1},
\end{equation}
which represents a classical state \eqref{classicalstates}. 
Thus, the set of classical spin-1 states can be identified with the
set of separable symmetric states of two qubits.  \\

This equivalence can also be shown indirectly using the PPT criterion. Indeed, there is a remarkable connection between the partial transpose of a state $\rho$ and the Bloch matrix $X$ of $\rho$. Namely, one can easily check that
\begin{equation}
\label{ppt bloch}
\rho^{\textrm{PT}}=\frac{1}{2} R X R^\dagger
\end{equation}
with the unitary matrix
\begin{equation}
R=\frac{1}{\sqrt{2}}\left(
\begin{array}{cccc}
 1 & 0 & 0 & 1 \\
 0 & 1 & -i & 0 \\
 0 & 1 & i & 0 \\
 1 & 0 & 0 & -1 \\
\end{array}
\right).
\end{equation}
Therefore the Bloch matrix is nothing but the partial transpose of
$\rho$ expressed in a different basis. As shown in section
\ref{section quantumness}, a necessary and sufficient condition for
classicality is that $X$ be positive semi-definite. As the eigenvalues
are unchanged by the change of basis \eqref{ppt bloch} (but for a
factor $\frac12$), this condition is equivalent to the positive
semi-definiteness of $\rho^{\textrm{PT}}$, which in turn is equivalent
to separability. In other words, this proves that a spin-1 state is
entangled (when seen as a bipartite system) if and only if its quantumness is non-zero. 
 
Any separable state can be written in the form \eqref{sep mixed states}, with possibly $\rho_i^{(1)} \neq \rho_i^{(2)}$. If that state lies in the subspace spanned by \eqref{dicke}, then necessarily, from the considerations above, $\rho_i^{(1)}=\rho_i^{(2)}$, so that $\rho$ can be cast in the form  
\begin{equation}
\label{entcoh}
\rho = \sum_i w_i \rho_i \otimes \rho_i,
\end{equation}
with $\rho_i$ spin-$\frac12$ coherent states, that is, a form where
the particle-exchange invariance appears also on the level of the
density matrix. 

\subsection{Quantumness and entanglement}
\label{Quantum-ent}
In Section \ref{purespin1} and \ref{mixed} we related quantumness of a
state $\rho$ to the smallest eigenvalue of its Bloch matrix
\eqref{Bloch matrix}. If this smallest eigenvalue is denoted by
$\lambda$, then from the correspondence \eqref{ppt bloch}, the
smallest eigenvalue of $\rho^{\textrm{PT}}$ is equal to
$\lambda/2$. In the case of a bipartition of two spin-$\frac12$
states,  $\rho^{\textrm{PT}}$ has at most one negative eigenvalue
\cite{Verstraete2001}, so that
negativity \eqref{negativity} reduces to $\mathcal{N}(\rho)
=-\lambda/2$. In the case of pure states, the
concurrence defined in \eqref{concurrence} reduces to  
\begin{equation}\label{Cl}
C(\ketbra\psi)=-\lambda.
\end{equation}
Of course, as is expected for pure states,  the negativity and the concurrence are simply related by $C(\ketbra\psi)=2\mathcal{N}(\ketbra\psi)$ \cite{Verstraete2001}. 


The function $f(\lambda)$ defined in \eqref{foflambda} thus allows us
to express quantumness as function of negativity for pure spin-1
states. For mixed spin-1 states Eq.~(\ref{upperboundmixed}) becomes
\begin{equation}
d_{HS}(\rho,\mathcal{C})\leq f(-2\mathcal{N}(\rho)),
\end{equation}
and as we showed equality holds for pure states.
Furthermore, it gives an insight into the geometry of
entangled states as it allows one to connect negativity to a geometric
property, namely  the Hilbert-Schmidt distance $d_{HS}$ from an entangled state
to the set $\mathcal{C}$ of symmetric separable states. In general,
since the closest separable state found in \cite{GeometryOfEntStates}
is non-symmetric, the corresponding minimal Hilbert-Schmidt distance
obtained in \cite{GeometryOfEntStates} is smaller than the one we get
as we consider the distance to symmetric separable states only.   

\section{Conclusion}
In this paper we investigated the quantumness of \mbox{spin-1} states,
defined as Hilbert-Schmidt distance to the convex set of classical
spin-1 states. We found the analytical solution for the quantumness
$Q(\ket\psi)$ of arbitrary pure states. It can be expressed as a function
of the smallest eigenvalue of the Bloch matrix associated with $\ket\psi$.
For mixed states, the same function appears to give an upper bound for
$Q(\rho)$ according to extensive numerical investigations. We
established the connection of $Q(\rho)$ with entanglement measures.

The closest classical state also provides a classicality witness, in the
spirit of \cite{BerKra09}. Our derivations provide another example of
the usefulness of the tensor representation of spin states \cite{PRL2015}. \\

{\bf Acknowledgments:} We thank the Deutsch-Franz\"osische
Hochschule (Universit\'e franco-allemande) for support, grant
number CT-45-14-II/2015.

\appendix

\section*{Appendix: Analytic calculation of the minima}
\label{Analytic calculation of the minima}

\begin{figure}[h]
\begin{center}
\begin{tikzpicture}[xscale=6.5, yscale=6.5]

    \draw [-,thick] (0,1) node (yaxis) [left] {}
            |- (1,0) node (xaxis) [below] {};
     
        \coordinate (diag_1) at (0,0);
    \coordinate (diag_2) at (1,0.6);
        \coordinate (b_2) at (1,0);
        \coordinate (orig) at (0,1);

        \coordinate (c) at (intersection of diag_1--diag_2 and orig--b_2);

        \draw [draw=none,fill=blue, fill opacity=0.3] (0,1) |- (1,0);
        \draw [draw=none,fill=red, fill opacity=0.3] (0,1) |- (c);

        \draw (orig) -- (b_2);

        \begin{scope}[font=\footnotesize]
            \foreach \y in {0,1}
                \draw (0,\y) node [left] {\y} --++(0.01,0);
            \foreach \x in {0,1}
                \draw (\x,0) node [below] {\x} --++(0.0,0.02);
        \end{scope}
      
   \node [below] at (0.5,0) {\LARGE $v$};
   \node [left] at (0,0.5) {\LARGE{$u$}};
  \node [right] at (0,0.7) {\large $(1)$};   
  \node [below] at (0.25,0.73) {\large $(2)$};
  \node [above] at (0.25,0.38) {\large $(3)$};       
  \node [right] at (0,0.2) {\large $(4)$};
  \node [above] at (0.4,0) {\large $(5)$};
  \node [below] at (0.78,0.2) {\large $(6)$};
    \draw[dashed] (yaxis |- c) node[left] (lambday) { $1-\lambda^2$}
            -| (xaxis -| c) node[below] (lambdax) { $\lambda^2$};
            
               \draw [-,thick] (0,1) node (yaxis) [left] {}
            |- (1,0) node (xaxis) [right] {};

    \end{tikzpicture}
\end{center}
\caption{Visualisation of the allowed parameter range of the variables
  $u$ and $v$. The upper area corresponds to $u\geq 1-\lambda^2$,
  while the lower corresponds to $u\leq 1-\lambda^2$. We call the
  function $F(u,v,g)$ restricted to the upper (lower) area $D$ ($E$). }  
\label{fig:lowerboundV6case1}
\end{figure}
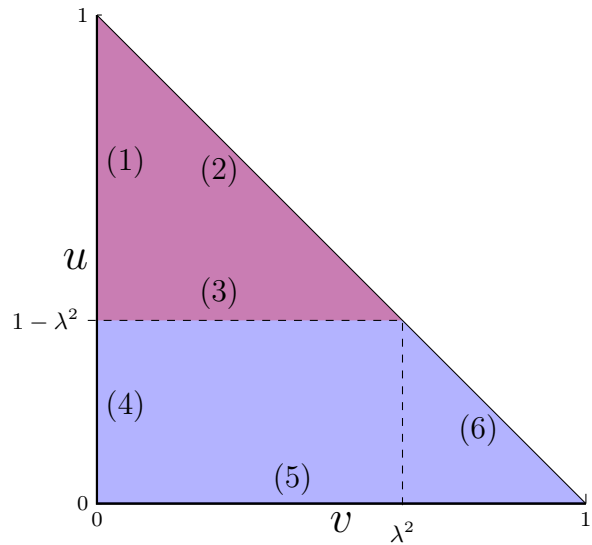

Here we will calculate the minimal value of $F$ defined in
\eqref{mingoal} under the constraints (\ref{para range}). If
$\lambda=0$, the minimum of $F$ is zero. We 
exclude this case in the following for convenience and restrict
ourselves to the interval $\lambda\in ]-1/2,0[$. 
We will use the fact that the minimal value of a function, restricted
to a certain parameter range, has its minimal value either on a
critical point or at the border of the parameter range. This will give
a list of candidates for the global minimum. The smallest value in
this list is then the global minimum.   

To calculate the minimal value, we distinguish two cases,
$u\geq1-\lambda^2$ and $u< 1-\lambda^2$.  In each case we can simplify the
problem by setting the variable $g$ to its optimal value. In the first
case $D:=F(g=\sqrt{1-\lambda^2})$ so that the third term vanishes, and
in the second case $E:=F(g=\sqrt{u})$, which makes the last term as
small as possible.   

In both cases the new functions 
\begin{align}
D&=(1-u)^2+(\lambda+v)^2, \\
E&=(1-u)^2+(\lambda+v)^2+ 2(\sqrt{1-\lambda^2}-\sqrt{u})^2
\end{align}
do not have critical points in the allowed parameter range of $u$ and $v$ \eqref{para range}, since $\nabla_{u,v} D=0$
is only solved by $(u,v)=(1,-\lambda)$, which is outside the parameter range for $\lambda<0$, and 
$\nabla_{u,v} E=0$ is only solved by $(u,v)=(\sqrt[3]{1-\lambda
  ^2},-\lambda)$, which is also outside the parameter range for
$\lambda<0$, since it contradicts the condition $u+v\le 1$. Therefore
both functions have to have 
their minimal value on the borders of the parameter range depicted in
Fig.~\ref{fig:lowerboundV6case1}. The function $D$ restricted to the
line (1) in Fig.~\ref{fig:lowerboundV6case1} will be referred to as $D^1$, analogues $D^2$, $D^3$. These
three functions do not have a critical point on the interior of their
respective parameter ranges, so the minimal value must be in all three
cases on one of the two vertices. Consider the candidates for the
minimal value for the function $D$, as 
\begin{align}
\begin{rcases} 
&D^1(u=1) 			\\
&D^2(v=0) 
\end{rcases}&= \lambda^2\\
\begin{rcases}
&D^1(u=1-\lambda^2)\\
&D^3(v=0)
\end{rcases}&=	\lambda ^4+\lambda ^2\\
\begin{rcases}
&D^2(v=\lambda^2)\\
&D^3(v=\lambda^2)
\end{rcases}&=\lambda ^2 \left(2 \lambda ^2+2 \lambda +1\right)\,.
\end{align}
Comparing these values the minimal value of $D$ is found to be
$\lambda ^2 \left(2 \lambda ^2+2 \lambda +1\right)$.

The minimum of the function $E$ will be calculated analogously. The
function on the line (3) in Fig.~\ref{fig:lowerboundV6case1} will be referred to as $E^3$, similar $E^4$
on line (4), and so forth. The function $E^3$ is the same as $D^3$ so
its minimal value is also $\lambda ^2 \left(2 \lambda ^2+2 \lambda
  +1\right)$. 
 
The function $E^4(u)=(1-u)^2+\lambda^2+ 2(\sqrt{1-\lambda^2}-\sqrt{u})^2$ has a critical value at $u=\sqrt[3]{1-\lambda ^2}$, which is larger than $1-\lambda^2$, and therefore outside the allowed range of the lower area in Fig.~\ref{fig:lowerboundV6case1}. So the minimal value is reached at the second of the two edges
\begin{align}
&E^4(u=0)=  3-\lambda ^2 , \\
&E^4(u=1-\lambda^2)= \lambda ^4+\lambda ^2  . 
\end{align}

The function $E^5(v)=1 + (v + \lambda)^2 + 2 (1 - \lambda^2)$ has a critical value in the allowed parameter range, at $v=-\lambda$, corresponding to a minimum 
\begin{align}
E^5(v=-\lambda)=3 - 2 \lambda^2.
\end{align}

%

The function $E^6(u)=(1-u)^2+ (\lambda+1-u)^2+2 \left(\sqrt{1-\lambda ^2} -\sqrt{u}\right)^2$ has a critical value in the allowed parameter range of $u \in [0,1-\lambda^2[$. The condition $\partial_u E^6=0$ gives  
\begin{equation}
1+\lambda +\frac{\sqrt{1-\lambda ^2}}{\sqrt{u}}-2 u=0, 
\end{equation}
with the substitution $u=y^2$ the optimal value of $u$ is given
through the real root $d$ of  
\begin{equation}
\sqrt{1-\lambda^2} +y(1+\lambda) -2 y^3=0,
\end{equation}
which is the same polynomial as in \eqref{PolynomThatGivesTheAngle}. The second derivative 
\begin{equation}
\frac{\partial^2 E^6}{\partial u^2}=\frac{\sqrt{1-\lambda ^2}}{u^{3/2}}+4
\end{equation}
is positive over the whole parameter range, so the critical point is a minimum, with value 
\begin{equation}
\label{d2 last case}
\ell(\lambda)=E^6(u=d^2).
\end{equation}
With this list of local minima, for all possible cases, the global minimum of \eqref{mingoal} is found to be \eqref{d2 last case}, which yields \eqref{l of lambda explicit definition}.

Proving Eq.~\eqref{l of lambda greater than f} is equivalent to showing
that $\ell(\lambda) \geq 
\frac{\lambda^2}{2}$. We have $E^6 \geq \frac{\lambda^2}{2}$, since 
\begin{equation}
\label{prove that l is larger than f}
E^6 -\frac{\lambda^2}{2} = 2 \left(\sqrt{u}-\sqrt{1-\lambda ^2}\right)^2+\frac{1}{2} \left[\lambda +2 (1-u)\right]^2,
\end{equation}
which, as the sum of squares, is always non-negative. Therefore the minimum of $E^6$ is also larger or equal to $\frac{\lambda^2}{2}$. As an immediate consequence, inequality \eqref{l of lambda greater than f} holds.

\end{document}